\documentclass{aa}
\usepackage{epsfig}
\usepackage{amsmath}
\usepackage{subfigure}
\usepackage{natbib}
\usepackage{multirow}
\usepackage{color}
\usepackage{longtable}
\usepackage{graphicx}
\usepackage{epstopdf}
\usepackage{booktabs}
\usepackage{txfonts}
\usepackage{graphicx}
\usepackage{txfonts}
\usepackage{color}
\usepackage{stfloats}
\usepackage{amsmath}
\usepackage{multirow}
\usepackage{epsfig}
\usepackage{amsmath}
\usepackage{subfigure}
\usepackage{natbib}
\usepackage{longtable}
\usepackage{graphicx}
\usepackage{epstopdf}
\usepackage{booktabs}
\usepackage{txfonts}
\usepackage[utf8]{inputenc}%
\usepackage{graphicx}
\usepackage{txfonts}
\usepackage{stfloats}

\newcommand{\kms}{km\,s$^{-1}$}

\newcommand{\nature}{Nature}

\newcommand{\prev}{Phys. Rev.}

\begin{document}

\title{Laboratory observation and astronomical search of 1-cyano propargyl radical, HCCCHCN \thanks{Based on observations carried out
with the Yebes 40m telescope (projects 19A003, 20A014, and 20D15). The 40m radiotelescope at Yebes Observatory is operated by the Spanish Geographic Institute
(IGN, Ministerio de Transportes, Movilidad y Agenda Urbana).}}

\author{
C.~Cabezas\inst{1},
M.~Nakajima\inst{2},
C.~H.~Chang\inst{3},
M.~Ag\'undez\inst{1},
Y.~Endo\inst{3},
and
J.~Cernicharo\inst{1}
}

\institute{Grupo de Astrof\'isica Molecular, Instituto de F\'isica Fundamental (IFF-CSIC), C/ Serrano 121, 28006 Madrid, Spain.
\email carlos.cabezas@csic.es
\and Department of Basic Science, Graduate School of Arts \& Sciences, The University of Tokyo, Komaba 3-8-1, Meguro-ku, Tokyo, 153-8902, Japan
\and Department of Applied Chemistry, Science Building II, National Yang Ming Chiao Tung University, 1001 Ta-Hsueh Rd., Hsinchu 30010, Taiwan
}

\date{Received; accepted}

\abstract
        {The reaction between carbon atoms and vinyl cyanide, CH$_2$CHCN, is a formation route to interstellar 3-cyano propargyl radical, CH$_2$C$_3$N, a species that has recently been discovered in space. The 1-cyano propargyl radical (HC$_3$HCN), an isomer of CH$_2$C$_3$N, is predicted to be produced in the same reaction at least twice more efficiently than CH$_2$C$_3$N. Hence, HC$_3$HCN is a plausible candidate to be observed in space as well.}
        {We aim to generate the HC$_3$HCN radical in the gas phase in order to investigate its rotational spectrum. The derived spectroscopic parameters for this species will be used to obtain reliable frequency predictions to support its detection in space.}
        {The HC$_3$HCN radical was produced by an electric discharge, and its rotational spectrum was characterized using a Balle-Flygare narrowband-type Fourier-transform microwave spectrometer operating in the frequency region of 4-40 GHz. The spectral analysis was supported by high-level ab initio calculations.}
        {A total of 193 hyperfine components that originated from 12 rotational transitions, $a$- and $b$-type, were measured for the HC$_3$HCN radical. The analysis allowed us to accurately determine 22 molecular constants, including rotational and centrifugal distortion constants as well as the fine and hyperfine constants. Transition frequency predictions were used to search for the HC$_3$HCN radical in TMC-1 using the QUIJOTE survey between 30.1-50.4 GHz. We do not detect HC$_3$HCN in TMC-1 and derive a 3$\sigma$ upper limit to its column density of 6.0$\times$10$^{11}$ cm$^{-2}$.}


\titlerunning{Laboratory detection of HC$_3$HCN}
\authorrunning{Cabezas et al.}

\maketitle

\section{Introduction}

The radioastronomical discovery of new molecules in space is highly dependent on the availability of precise rotational spectroscopic laboratory data. Only a few molecular species have been detected in space prior to their characterization in the laboratory. Some examples of this type are HCO$^+$ \citep{Buhl1970}, C$_4$H \citep{Guelin1978}, HCS$^+$ \citep{Thaddeus1981}, C$_5$H \citep{Cernicharo1986}, C$_6$H \citep{Suzuki1986}, and C$_3$H$^+$ \citep{Pety2012}. All of them were confirmed later on in the laboratory. Other cases such as C$_5$N$^-$ \citep{Cernicharo2008}, MgC$_3$N and MgC$_4$H \citep{Cernicharo2019}, HC$_5$NH$^+$ \citep{Marcelino2020}, MgC$_5$N and MgC$_6$H \citep{Pardo2021}, and H$_2$NC \citep{Cabezas2021a} have not been yet observed in the laboratory. Although most of these molecules are open-shell species, they display almost trivial spectral patterns, except for H$_2$NC. These patterns allowed their identifications in the line surveys only supported by dedicated ab initio calculations. However, other interstellar molecules and potential candidates have much more complex rotational spectra due to the presence of two or three nuclei with nonzero nuclear spins, which causes complicated line splitting by interactions of different types of angular momenta. In these cases, ab initio calculations are not enough to identify their spectral features in the line surveys, and the laboratory experiments are invaluable tools \citep{Cabezas2021c}.

An example of this type of open-shell species is the 3-cyano propargyl radical, CH$_2$C$_3$N, which was recently discovered in space \citep{Cabezas2021b}. Its astronomical identification was possible through high-resolution laboratory experiments \citep{Chen1998, Tang2001}, which provided frequency predictions with an accuracy better than 10\,kHz in the 30-50 GHz region. The hyperfine components for seven rotational transitions of CH$_2$C$_3$N were detected in the cold dark cloud TMC-1 using the Yebes 40m telescope. The matching of these precise frequencies to interstellar features was then aided by the sharpness of the radio emission features in TMC-1, around 40 kHz at 40 GHz.

One of the routes to form CH$_2$C$_3$N in space is the reaction C + CH$_2$CHCN $\rightarrow$ CH$_2$C$_3$N + H. This reaction has been studied using crossed molecular beam experiments and theoretical calculations \citep{Su2005,Guo2006}. These studies indicate that the reaction is barrierless and occurs through H atom elimination, yielding as main products the radicals 3-cyano propargyl (CH$_2$C$_3$N) and its isomer 1-cyano propargyl (HC$_3$HCN). The latter is inferred to be produced at least twice more efficiently than the former \citep{Guo2006}, and thus, it is a good candidate for detection in TMC-1 as well.

As far as we could ascertain from the literature, no spectroscopic work on 1-cyano propargyl radical (HC$_3$HCN) has been published so far. The only information about this species is that mentioned previously about crossed molecular beam experiments and theoretical calculations \citep{Su2005,Guo2006}. In the present work, we report the first rotational study of HC$_3$HCN using Fourier transform microwave (FTMW) spectroscopy supported by ab initio calculations. In addition, we carried out an astronomical search of HC$_3$HCN in TMC-1 using accurate frequency predictions derived from our spectroscopic study.

\begin{figure}
\centering
\includegraphics[angle=0,width=0.45\textwidth]{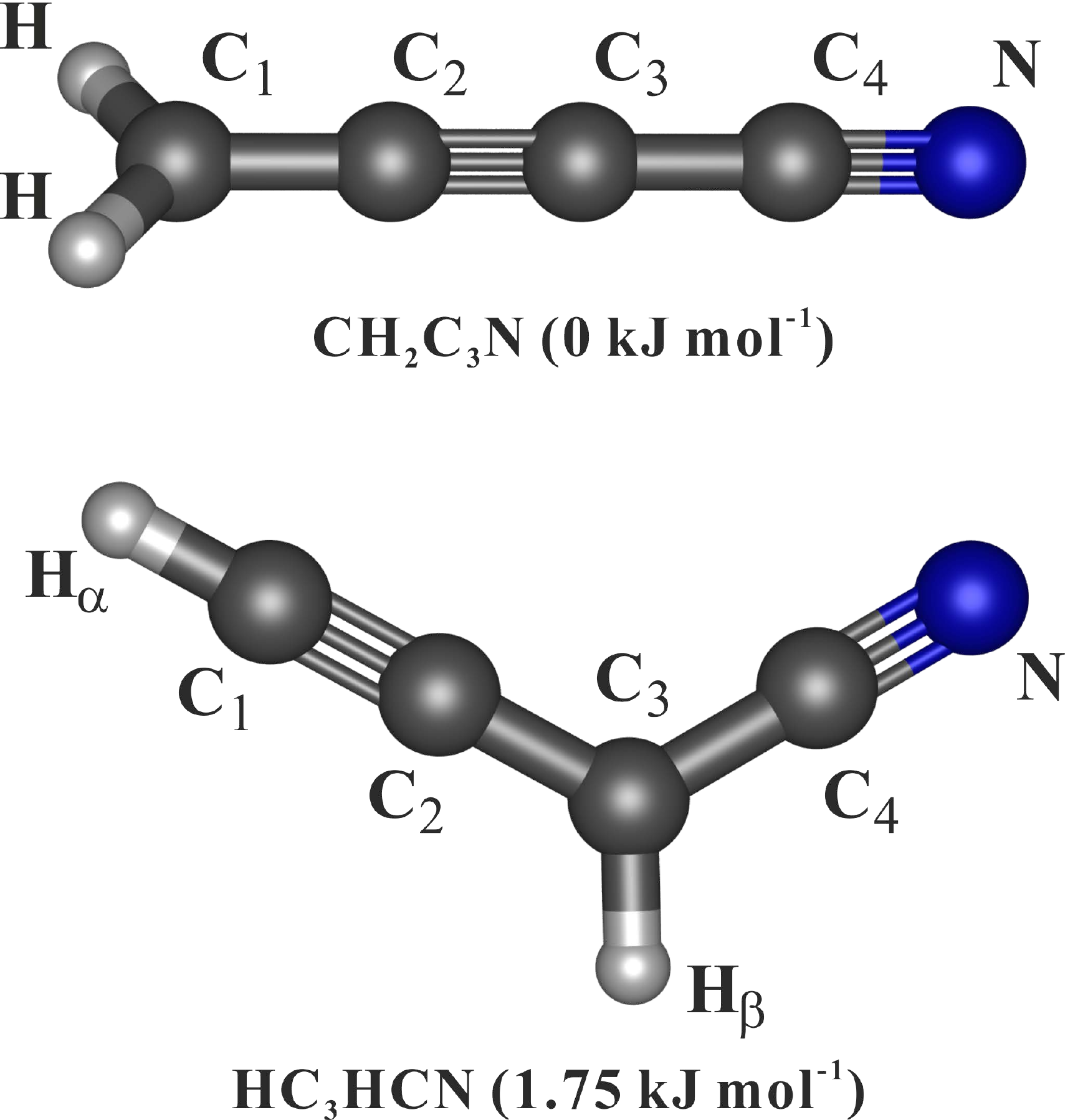}
\caption{Molecular structures and relative energies of the CH$_2$C$_3$N and HC$_3$HCN radicals.} \label{isomers}
\end{figure}

\section{Experiment}
\label{exp}

The rotational spectrum of the HC$_3$HCN radical was observed using a Balle-Flygare narrowband-type FTMW spectrometer operating in the frequency region of 4-40 GHz \citep{Endo1994,Cabezas2016}. The short-lived species HC$_3$HCN was produced in a supersonic expansion by a pulsed electric discharge of a gas mixture of CH$_2$CHCN (0.2\%) and C$_2$H$_2$ (0.4\%) diluted in Ar. This gas mixture was flowed through a pulsed-solenoid valve that is accommodated in the backside of one of the cavity mirrors and aligned parallel to the optical axis of the resonator. A pulse voltage of 900 V with a duration of 450 $\mu$s was applied between stainless-steel electrodes attached to the exit of the pulsed discharge nozzle (PDN), resulting in an electric discharge synchronized with the gas expansion. The resulting products generated in the discharge were supersonically expanded, rapidly cooled to a rotational temperature of $\sim$2.5\,K between the two mirrors of the Fabry-P\'erot resonator, and then probed by FTMW spectroscopy. For measurements of the paramagnetic lines, the Earth’s magnetic field was cancelled by using three sets of Helmholtz coils placed perpendicularly to one another. Since the PDN is arranged parallel to the cavity of the spectrometer, it is possible to suppress the Doppler broadening of the spectral lines, which allows resolving small hyperfine splittings. The spectral resolution is 5\,kHz, and the frequency measurements have an estimated accuracy better than 3\,kHz.

\section{Ab initio calculations}
\label{calc}

\begin{table}
\small
\caption{Spectroscopic parameters of HC$_3$HCN (all in MHz).}
\label{constants}
\centering
\begin{tabular}{lcc}
\hline \hline
\multicolumn{1}{c}{Parameter}  & \multicolumn{1}{c}{Experimental} & \multicolumn{1}{c}{Theoretical} \\
\hline
$A$                          &  ~  29173.01440(72)\,$^a$    &  ~    27909\,$^b$       \\
$B$                          &  ~      2807.94534(25)       &  ~     2835\,$^b$       \\
$C$                          &  ~      2556.92917(25)       &  ~     2574\,$^b$       \\
$\Delta_N$                   &  ~      0.0014120(19)        &  ~    0.0014\,$^c$      \\
$\Delta_{NK}$                &  ~     $-$0.109826(50)       &  ~    $-$0.091\,$^c$    \\
$\delta_N$                   &  ~      0.0003574(45)        &  ~    0.000345\,$^c$    \\
$\varepsilon_{aa}$           &  ~      $-$60.5224(30)       &  ~    $-$51.35\,$^c$    \\
$\varepsilon_{bb}$           &  ~      $-$5.3817(12)        &  ~    $-$6.385\,$^c$    \\
$\varepsilon_{cc}$           &  ~      $-$0.5405(13)        &  ~    $-$1.60\,$^c$     \\
$\varepsilon_{ab}$           &  ~      $\mp$1.214(73)$^d$   &  ~  $\mp$1.64\,$^c$     \\
$a_F$$^{\rm(H_\alpha)}$      &  ~      $-$32.2969(32)       &  ~   $-$31.70\,$^e$     \\
$T_{aa}$$^{\rm(H_\alpha)}$   &  ~        8.7599(36)         &  ~       9.14\,$^e$     \\
$T_{bb}$$^{\rm(H_\alpha)}$   &  ~      $-$7.9786(61)        &  ~    $-$8.43\,$^e$     \\
$T_{ab}$$^{\rm(H_\alpha)}$   &  ~      $\pm$12.58(20)$^d$   &  ~ $\pm$13.55\,$^e$     \\
$a_F$$^{\rm(H_\beta)}$       &  ~      $-$51.3475(46)       &  ~   $-$55.10\,$^e$     \\
$T_{aa}$$^{\rm(H_\beta)}$    &  ~      $-$25.3318(27)       &  ~   $-$29.31\,$^e$     \\
$T_{bb}$$^{\rm(H_\beta)}$    &  ~       24.5181(48)         &  ~      28.42\,$^e$     \\
$T_{ab}$$^{\rm(H_\beta)}$    &  ~     [$\pm$0.79]$^f$       &  ~  $\pm$0.79\,$^e$     \\
$a_F$$^{\rm(N)}$             &  ~        6.6999(27)         &  ~     1.1852\,$^e$     \\
$T_{aa}$$^{\rm(N)}$          &  ~      $-$10.3977(22)       &  ~   $-$11.33\,$^e$     \\
$T_{bb}$$^{\rm(N)}$          &  ~      $-$9.1827(48)        &  ~    $-$9.51\,$^e$     \\
$T_{ab}$$^{\rm(N)}$          &  ~       [$\pm$1.72]         &  ~  $\pm$1.72\,$^e$     \\
$\chi_{aa}$$^{\rm(N)}$       &  ~      $-$2.7743(22)        &  ~    $-$2.82\,$^e$     \\
$\chi_{bb}$$^{\rm(N)}$       &  ~        0.6098(42)         &  ~       0.57\,$^e$     \\
$\chi_{ab}$$^{\rm(N)}$       &  ~     [$\pm$2.70]           &  ~  $\pm$2.70\,$^e$     \\
\hline
$|\mu_a|$                    &                              &  ~   3.70\,$^{b,g}$     \\
$|\mu_b|$                    &                              &  ~   1.77\,$^{b,g}$     \\
\hline
\end{tabular}
\tablefoot{
\tablefoottext{a}{Numbers in parentheses are 1$\sigma$ uncertainties in units of the last digits.}
\tablefoottext{b}{RCCSD(T)/cc-pCVQZ.}
\tablefoottext{c}{MP2/cc-pVTZ.}
\tablefoottext{d}{Only the relative signs are determined.}
\tablefoottext{e}{QCISD/cc-pVTZ.}
\tablefoottext{f}{Values in brackets were fixed to the theoretical values.}
\tablefoottext{g}{In units of Debye.}
}
\end{table}

The geometry optimization calculation of HC$_3$HCN was carried out using the spin-restricted coupled cluster method with single, double, and perturbative triple excitations (RCCSD(T); \citealt{Raghavachari1989}) with all electrons (valence and core) correlated and the Dunning correlation-consistent basis sets with polarized core-valence
correlation quadruple-$\zeta$ (cc-pCVQZ; \citealt{Woon1995}). At the optimized geometry, electric dipole moment components along the $a$- and $b$-inertial axes were calculated at the same level of theory as that for the geometrical optimization. The calculations were performed using the MOLPRO 2020.2 program \citep{Werner2020}. The fine and hyperfine
coupling constants of the HC$_3$HCN radical were estimated at the second-order M{\o}ller-Plesset perturbation (MP2; \citealt{Moller1934}) and the quadratic configuration interaction with single- and double-excitation (QCISD; \citealt{Pople1987}) levels of calculations, with the cc-pVTZ basis set \citep{Woon1995}. Harmonic frequencies were computed at the MP2/cc-pVTZ level of theory to estimate the centrifugal distortion constants. These calculations were performed using the Gaussian16 \citep{Frisch2016} program package. The calculated geometry of HC$_3$HCN in the $^{2}A''$ ground electronic state is depicted in Fig. \ref{isomers} together with that for the CH$_2$C$_3$N isomer. The calculated molecular parameters and the dipole moment components are summarized in Table \ref{constants}.

In addition, we carried out optimization calculations for the CH$_2$C$_3$N isomer in order to ascertain the relative energies of this species and HC$_3$HCN. The RCCSD(T)/cc-pCVQZ energies with the zeropoint energy corrections at RCCSD/cc-VQZ level of theory indicate that the CH$_2$C$_3$N isomer is slightly more stable than HC$_3$HCN by 1.75 kJ mol$^{-1}$. \citet{Su2005} carried out quantum chemical calculations for HC$_3$HCN and CH$_2$C$_3$N isomers at the B3LYP/6-311G(d,p)//CCSD(T)/6-311G(d,p) level of theory. They found that CH$_2$C$_3$N is more stable by 14.9 kJ mol$^{-1}$ when B3LYP/6-311G(d,p) energies are taken into account. However, HC$_3$HCN is 2.1 kJ mol$^{-1}$ more stable than CH$_2$C$_3$N if the CCSD(T)/6-311G(d,p) energies are considered. Given the small energy differences, which are of the order of the error of the calculations, we can consider that the two isomers CH$_2$C$_3$N and HC$_3$HCN are nearly isoenergetic.

\section{Rotational spectrum analysis}
\label{rot_spe}

\begin{figure*}
\centering
\includegraphics[angle=0,width=0.92\textwidth]{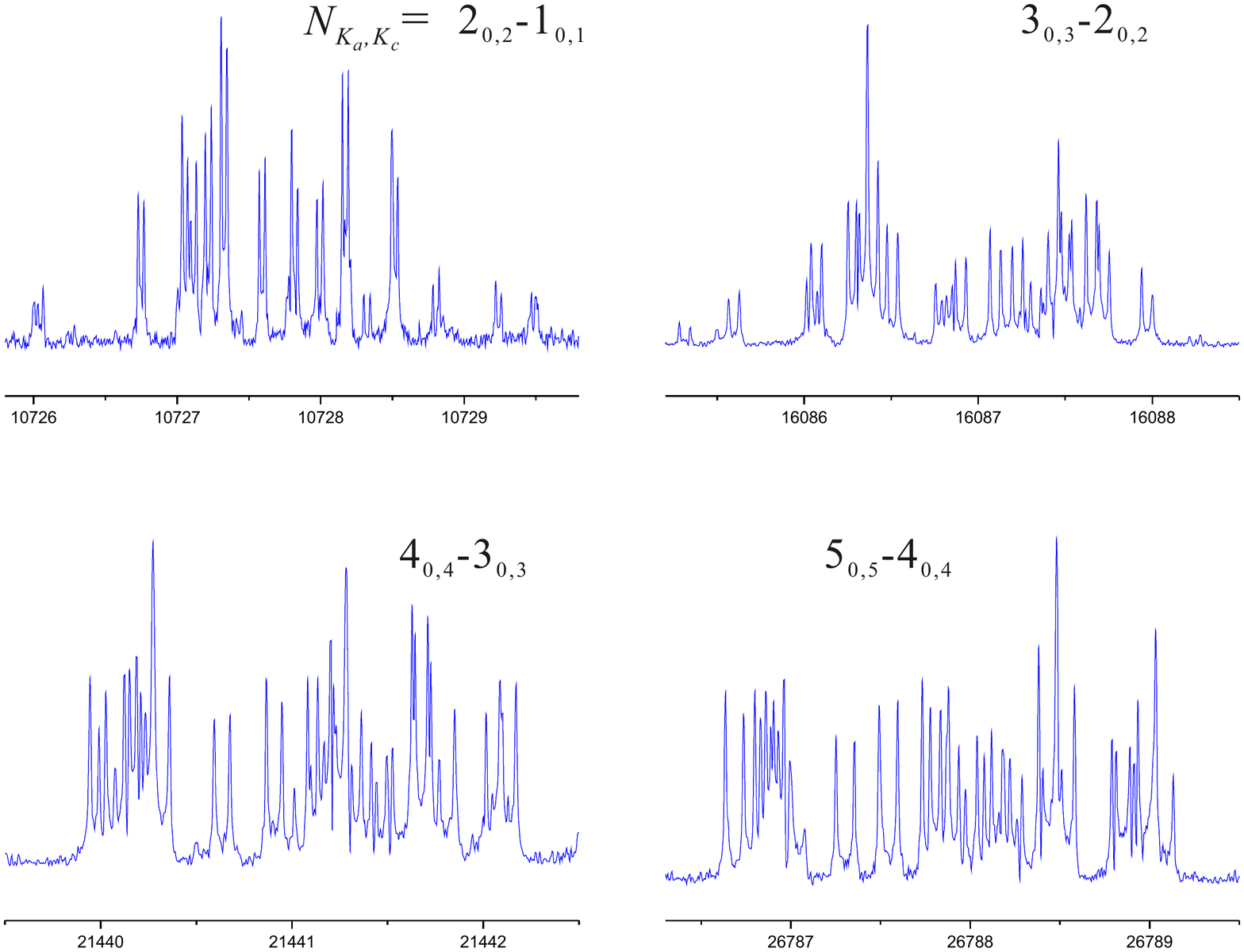}
\caption{Four $K_{a}$ = 0 $a$-type rotational transitions for the HC$_3$HCN radical. For the 2$_{0,2}$-1$_{0,1}$, 3$_{0,3}$-2$_{0,2}$ , and 4$_{0,4}$-3$_{0,3}$ transitions, additional hyperfine components, which are not shown in this figure, were also observed.  The abscissa corresponds to the frequency of the lines in MHz. The spectra were achieved by 100 shots of accumulation and a step scan of 0.5 MHz with a repetition rate of 10 Hz. The coaxial arrangement of the adiabatic expansion and the resonator axis produces an instrumental Doppler doubling.
The resonance frequencies are calculated as the average of the two Doppler components.} \label{ftmw}
\end{figure*}

The predicted $a$-type dipole moment for the HC$_3$HCN radical is 3.70\,D, and thus, the $a$-type rotational transitions are expected to be fairly strong. Hence, we first tried to observe this type of transition. In particular, we searched for the $N_{K_{a,}K_{c}}$= 2$_{0,2}$-1$_{0,1}$ transition, which is predicted around 10.8 GHz. A group of more than 20 paramagnetic lines were observed in the 10.7 GHz region. The 3$_{0,3}$-2$_{0,2}$, 4$_{0,4}$-3$_{0,3}$ , and 5$_{0,5}$-4$_{0,4}$ transitions were then searched around 16.1, 21.5, and 26.8 GHz. Three groups of paramagnetic lines with a similar spectral pattern to that observed in the 10.7 GHz region were found around these frequencies. Figure \ref{ftmw} shows the recorded spectra for the $a$-type $K_{a}$ = 0 rotational transitions. The HC$_3$HCN radical was readily confirmed as the spectral carrier based on the following arguments. (i) The observed transition frequencies agree well with the calculated frequencies, (ii) each transition has a hyperfine spectral structure similar to that expected for an open-shell species with three coupling nuclei, and (iii) the lines exhibit the paramagnetic behavior.

The hyperfine components for these four transitions were analyzed using an appropriate Hamiltonian for an asymmetric top molecule with a doublet electronic state ($^2A^{''}$) and the centrifugal distortion expressed in Watson’s $A$-reduced form \citep{Watson1967}. The employed Hamiltonian has the following form:
\begin{equation}
\textbf{H} = \textbf{H}_{rot} + \textbf{H}_{sr} + \textbf{H}_{mhf} + \textbf{H}_Q
,\end{equation}
where \textbf{H}$_{rot}$ contains rotational and centrifugal distortion parameters, $\textbf{H}_{sr}$ is the spin-rotation term, $\textbf{H}_{mhf}$ represents the magnetic hyperfine coupling interaction term due to the hydrogen and nitrogen nuclei, and $\textbf{H}_Q$ represents the nuclear electric quadrupole interaction due to the nitrogen nucleus. A complete description of these terms can be found in \citet{Hirota1985}. The coupling scheme used is \textbf{J}\,=\,\textbf{N}\,+\,\textbf{S}, \textbf{F}$_1$\,=\,\textbf{J}\,+\,\textbf{I}$_1$, \textbf{F}$_2$\,=\,\textbf{F}$_1$\,+\,\textbf{I}$_2$, and \textbf{F}\,=\,\textbf{F}$_2$\,+\,\textbf{I}$_3$, where \textbf{I}$_1$\,=\,\textbf{I}(H$_{\beta}$), \textbf{I}$_2$\,=\,\textbf{I}(H$_{\alpha}$)\ and \textbf{I}$_3$\,=\,\textbf{I}(N).

An initial fit to the $K_{a}$ = 0 transition frequencies provided a first set of experimental constants for HC$_3$HCN, which was later on refined. The following assignment was performed in a classical bootstrap manner, where assigned transitions were used to improve the frequency predictions and to search for new ones. In this manner, we measured six $K_{a}$ = 1 $a$-type transitions with $N$ = 2, 3, and 4 and two $b$-type $Q$-branch transitions, 1$_{1,0}$-1$_{0,1}$ and 2$_{1,1}$-2$_{0,2}$. The final dataset consists of 193 hyperfine components that originated from 12 rotational transitions. Table \ref{tab_lab_freq} contains the experimental and calculated frequencies for all the observed hyperfine components as  well as their relative intensities within a particular rotational transition, normalized to the degeneracy of the lower level. Twenty-two molecular constants were determined by the least-squares analysis for all the observed transition frequencies. The standard deviation of the fit is 2.0 kHz, which is slightly smaller than the experimental accuracy of the measurements, indicating that the complicated hyperfine structures caused by the three nuclei are well described by the employed Hamiltonian. The determined molecular constants are summarized in Table~\ref{constants} along with those predicted by ab initio calculations.

\section{Discussion}
\label{discus}

As mentioned before, a total of 22 molecular constants were determined from the fit. For 2 of them, $\varepsilon_{ab}$ and $T_{ab}$$^{\rm(H_\alpha)}$, only the relative signs could be determined. In addition to these 22 parameters, we included in the fit 3 parameters, $T_{ab}$$^{\rm(H_\beta)}$, $T_{ab}$$^{\rm(N)}$ , and $\chi_{ab}$$^{\rm(N)}$, fixed to the theoretical values. This resulted in an improvement of the standard deviation of the fit from 2.4 kHz to the final deviation of 2.0 kHz. In general, Table \ref{constants} shows that the experimentally determined values agree well with the ab initio ones, which implies that the calculated molecular structure is reasonable. The $B$ and $C$ calculated constants show relative errors from the experimental values of 0.7 and 0.9 \%, respectively. In contrast, the relative error for the $A$ constant is fairly large, around 4.3\%. We tried to reproduce the experimental value of the $A$ constant by increasing the level of calculation, but no better results were found. We also considered that this discrepancy could be due to the vibrational contribution. However, our vibration-rotation calculations showed that this contribution is not that large, it is smaller than 0.5\%.

\begin{table}
\small
\caption{Spin-rotation interaction, Fermi coupling, and dipole-dipole constants for HC$_3$HCN and CH$_2$C$_3$N (all in MHz).}
\label{fine_hyper}
\centering
\begin{tabular}{lc|c}
\hline \hline
\multicolumn{1}{c}{Parameter}  & \multicolumn{1}{c}{HC$_3$HCN} & \multicolumn{1}{c}{CH$_2$C$_3$N} \\
\hline
$\varepsilon_{aa}$           &  ~       $-$60.5224       &  ~       $-$660.330       \\
$\varepsilon_{bb}$           &  ~        $-$5.3817       &  ~       $-$4.2396        \\
$\varepsilon_{cc}$           &  ~        $-$0.5405       &  ~       $-$0.3955        \\
$a_F$$^{\rm(H_\alpha)}$      &  ~       $-$32.2969       &  ~       $-$50.69\,$^a$   \\
$a_F$$^{\rm(H_\beta)}$       &  ~       $-$51.3475       &  ~       $-$50.69\,$^a$   \\
$a_F$$^{\rm(N)}$             &  ~           6.6999       &  ~        5.1618          \\
$T_{aa}$$^{\rm(N)}$          &  ~       $-$10.3977       &  ~       $-$8.2149        \\
$T_{bb}$$^{\rm(N)}$          &  ~        $-$9.1827       &  ~       $-$5.362         \\
\hline
\hline
\end{tabular}
\tablefoot{
\tablefoottext{a}{The notation ${\rm(H_\alpha)}$ and ${\rm(H_\beta)}$ does not apply to CH$_2$C$_3$N. }
}
\end{table}

\begin{figure}
\centering
\includegraphics[angle=0,width=0.4\textwidth]{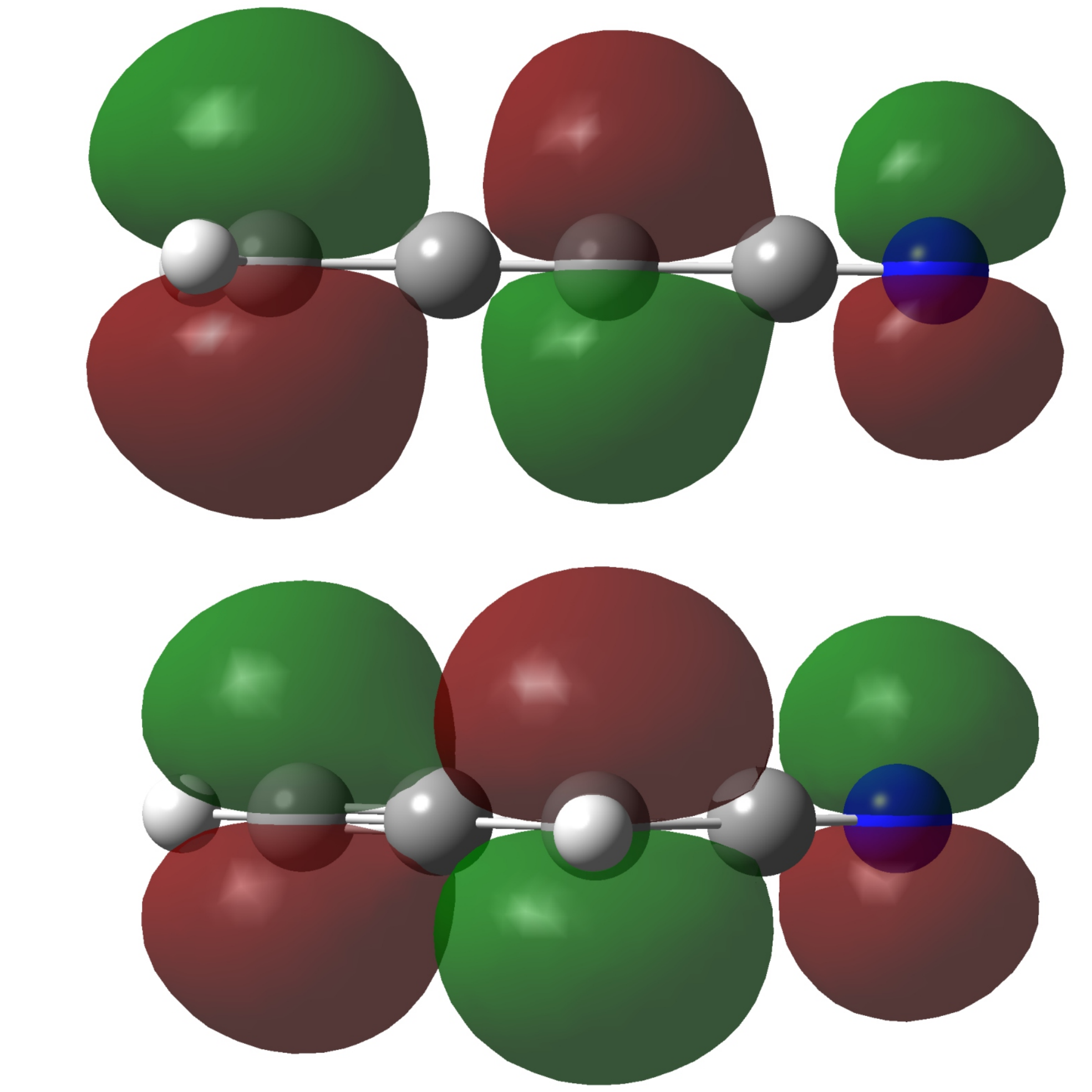}
\caption{Molecular orbital for the unpaired electron in CH$_2$C$_3$N (top) and HC$_3$HCN (bottom).} \label{fermi}
\end{figure}

If scaled by the rotational constants, the spin-rotation interaction constants determined for HC$_3$HCN are similar to those for the CH$_2$C$_3$N isomer, as shown in Table \ref{fine_hyper}.The $A$, $B$ , and $C$ constants for CH$_2$C$_3$N are 288000, 2195.08395, and 2177.77590 MHz, respectively \citep{Tang2001}. This implies that the two radicals have a nearly equal-energy excited states, $^2B^{2}$ in the case of CH$_2$C$_3$N and $^2A^{'}$ for HC$_3$HCN, which contribute mainly to the spin-rotation interaction constants in the ground state through the spin-orbit coupling. The Fermi coupling constants give direct information of the unpaired electron density on the nuclei. The $a_F$$^{\rm(H_\alpha)}$ and $a_F$$^{\rm(H_\beta)}$ have negative values due to the spin-polarization in the C-H bonds. The difference in the absolute values between $a_F$$^{\rm(H_\alpha)}$ and $a_F$$^{\rm(H_\beta)}$ indicate that the unpaired electron density in C$_1$ is lower than that in C$_3$. This is consistent with the predictions of the molecular structure and the unpaired electron orbital depicted in Figure \ref{fermi}.

As shown in Table \ref{fine_hyper}, the $a_F$$^{\rm(H_\beta)}$ value is very similar to that found for the hydrogen nuclei of CH$_2$C$_3$N. Hence, it can be inferred that the unpaired electron density in C$_1$ of CH$_2$C$_3$N is very much like that in C$_3$ of HC$_3$HCN, which is illustrated in Figure \ref{fermi}. As for the nitrogen nucleus, the $a_F$ constant is almost similar in HC$_3$HCN and CH$_2$C$_3$N. On the other hand, the determined $T_{aa}$ and $T_{bb}$ constants for the nitrogen nucleus in HC$_3$HCN are very similar and roughly satisfy the relation $T_{cc}$ = -2$T_{aa}$ = -2$T_{bb}$. This indicates that the orientation of the unpaired electron is along the $c$-axis and occupies a $p_{\pi}$ molecular orbital extending perpendicular to the molecular plane. This is shown in Figure \ref{fermi}. The values for $T_{aa}$ and $T_{bb}$ in both HC$_3$HCN and CH$_2$C$_3$N radicals are comparable, which indicates that the unpaired electron density around the nitrogen nucleus in both radicals is almost the same. All these facts reveal similarities between the ground states of both radicals.

Three canonical forms can be used to describe the HC$_3$HCN radical: H--C$\equiv$C--$\dot{\rm C}$H--C$\equiv$N, H--$\dot{\rm C}$=C=CH--C$\equiv$N, and H--C$\equiv$C--CH=C=$\dot{\rm N}$. The calculated bond lengths for HC$_3$HCN are 1.217 $\mathring{\rm A}$, 1.385 $\mathring{\rm A}$, 1.407 $\mathring{\rm A}$, and 1.164 $\mathring{\rm A}$ for C$_1$--C$_2$, C$_2$--C$_3$, C$_3$--C$_4$ , and C$_4$--N, respectively. The bond distances C$_1$--C$_2$ and C$_4$--N are almost identical to those found in the molecule HC$_3$N \citep{Botschwina1993}, 1.206 $\mathring{\rm A}$ and 1.161 $\mathring{\rm A}$ for C--C and C--N triple bonds, respectively. The C$_2$--C$_3$ and C$_3$--C$_4$ lengths are slightly larger than the C--C single-bond distance in HC$_3$N. Although the three Lewis structures are contributing to the electronic structure for the $^2A^{''}$ ground state, the ab initio geometry indicates that the H--C$\equiv$C--$\dot{\rm C}$H--C$\equiv$N form contributes slightly more than the other two.
This agrees with the molecular orbital of the unpaired electron for HC$_3$HCN shown in Figure \ref{fermi} and the values determined for the Fermi constants.

\section{Astronomical search}
\label{astro}

From the spectroscopic parameters presented in Table \ref{constants}, we generated reliable frequency predictions to guide the astronomical search for HC$_3$HCN. The predictions have an accuracy better than 10 kHz in the Q band (30-50\,GHz) and $\sim$10-20 kHz in the W band (72-116\,GHz). The rotational partition functions we used in these predictions are listed in Table \ref{pfunction}. They were calculated considering a maximum value of $N$ = 40. The frequency predictions for HC$_3$HCN up to 300\,GHz calculated at T = 10\,K and 300\,K are available at the CDS, Tables A.2 and A.3 respectively.

\begin{table}
\begin{center}
\caption[]{Rotational partition function for HC$_3$HCN at different temperatures.}
\scalebox{1}{
\label{pfunction}
\begin{tabular}{cc}
\hline
\hline
Temperature/K   &\hfill $Q_r$\hfill\mbox{}\\
\hline
 300.000  & 694460.4     \\
 225.000  & 559886.3     \\
 150.000  & 388347.4     \\
  75.000  & 175077.8     \\
  37.500  &  66626.2     \\
  18.750  &  24000.2     \\
  10.000  &   9495.6     \\
   9.375  &   8634.1     \\
\hline
\end{tabular}
}
\end{center}
\end{table}

\begin{figure}
\centering
\includegraphics[angle=0,width=0.5\textwidth]{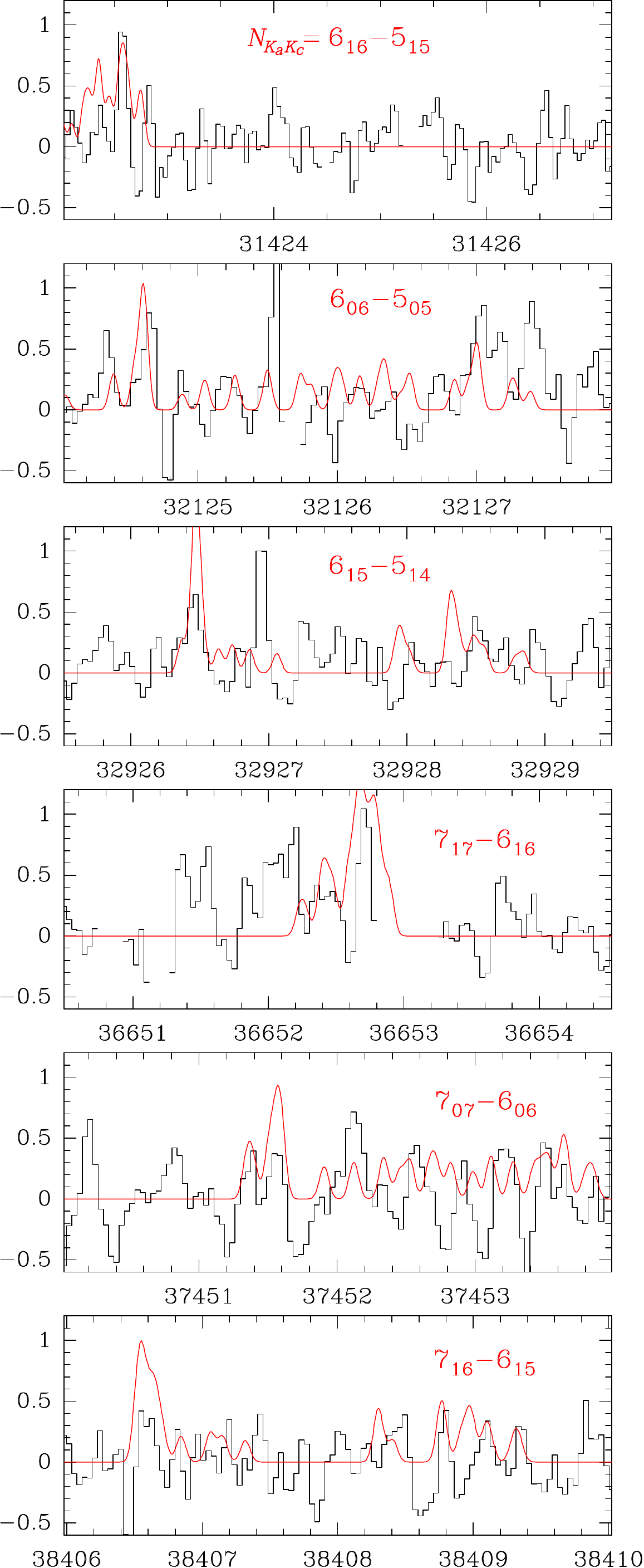}
\caption{Observed data of TMC-1 from the QUIJOTE line survey (black histogram) and synthetic spectra calculated adopting as column density the upper limit given in the text (red curve). For each rotational transition, the most intense hyperfine components are shown. The abscissa corresponds to the rest frequency assuming a local standard of rest velocity of 5.83\,\kms. The ordinate is the antenna temperature in millikelvin.} \label{tmc}
\end{figure}

We searched for the HC$_3$HCN radical toward the cold dark molecular cloud TMC-1, where the CH$_2$C$_3$N isomer has been detected \citep{Cabezas2021b}. The spectral data employed in this work are part of the QUIJOTE\footnote{\textbf{Q}-band \textbf{U}ltrasensitive \textbf{I}nspection \textbf{J}ourney to the \textbf{O}bscure \textbf{T}MC-1 \textbf{E}nvironment} line survey \citep{Cernicharo2021a}, performed toward TMC-1 ($\alpha_{J2000}=4^{\rm h} 41^{\rm  m} 41.9^{\rm s}$ and $\delta_{J2000}=+25^\circ 41' 27.0''$) in the Q band using the Yebes 40m radiotelescope. The observations were performed in three different observing runs during November 2019 and April 2021. The first observing run allowed the detection of the C$_3$N$^-$ and C$_5$N$^-$ anions \citep{Cernicharo2020a} and the discoveries of HC$_4$NC (\citet{Cernicharo2020b}), HC$_3$O$^+$ (\citet{Cernicharo2020c}), and HC$_5$NH$^+$ \citep{Marcelino2020}. Sensitivity improvements from additional observations performed in October 2020 resulted in the detection of HDCCN \citep{Cabezas2021c}, HC$_3$S$^+$ \citep{Cernicharo2021b}, and CH$_3$CO$^+$ \citep{Cernicharo2021b} and three isomers with the formula C$_4$H$_3$N \citep{Marcelino2021}. The last observing run performed between December 2020 and April 2021 resulted in new discoveries, including hydrocarbon species such as the CH$_2$CCH radical \citep{Agundez2021a}, CH$_2$CHCCH and CH$_2$CCHCCH \citep{Cernicharo2021d,Cernicharo2021e}, $l$-H$_2$C$_5$ \citep{Cabezas2021d}, $c$-C$_3$HCCH, $c$-C$_5$H$_6$ , and $c$-C$_9$H$_8$ \citep{Cernicharo2021f}; sulphur-bearing species such as NCS, HCCS, H$_2$CCS, H$_2$CCCS, C$_4$S, HCSCN, and HCSCCH \citep{Cernicharo2021g,Cernicharo2021h}; O-bearing complex organic molecules such as C$_2$H$_3$CHO, C$_2$H$_3$OH, HCOOCH$_3$ , and CH$_3$OCH$_3$ \citep{Agundez2021b}, and deuterated isotopologs such as CH$_2$DC$_3$N \citep{Cabezas2021e}.

All observations were carried out using the frequency-switching technique, with a frequency throw of 10\,MHz during the first two observing runs and a throw of 8\,MHz in the later ones. The intensity scale, the antenna temperature $T_A^*$, was calibrated using two absorbers at different temperatures and the atmospheric transmission model (ATM; \citealt{Cernicharo1985, Pardo2001}). Different frequency coverages were observed, 31.08-49.52 GHz and 31.98-50.42 GHz, which permit verifying that no spurious ghosts are produced in the down-conversion chain in which the signal coming from the receiver is downconverted to 1-19.5 GHz and then split into eight bands with a coverage of 2.5 GHz, each of which are analyzed by the FFTs. Calibration uncertainties were adopted to be 10~\% based on the observed repeatability of the line intensities between different observing runs. All data were analyzed using the GILDAS package\footnote{\texttt{http://www.iram.fr/IRAMFR/GILDAS}}.

The frequency predictions were implemented in the MADEX code \citep{Cernicharo2012} to compute synthetic spectra assuming local thermodynamic equilibrium. We used the dipole moment components from Table~\ref{constants} and assumed a rotational temperature of 7\,K \citep{Cabezas2021b} and a $\nu_{\rm LSR}$ = 5.83 km s$^{-1}$ \citep{Cernicharo2020a}. Figure \ref{tmc} shows the spectrum of TMC-1 at the frequencies of selected transitions of HC$_3$HCN, together with the synthetic spectra calculated with MADEX. Lines of HC$_3$HCN are not clearly seen in our TMC-1 data. The sensitivity of the QUIJOTE line survey varies between 0.17 and 0.3\,mK in the 31-50\,GHz domain. Adopting the observed 3$\sigma$ limits to the intensity of strongest components of HC$_3$HCN, we derive a 3$\sigma$ upper limit to the column density of this species in TMC-1 of 6.0$\times$10$^{11}$ cm$^{-2}$.

Given the column density of CH$_2$C$_3$N in TMC-1, 1.6\,$\times$\,10$^{11}$ cm$^{-2}$ \citep{Cabezas2021b}, the column density ratio between the two isomers HC$_3$HCN/CH$_2$C$_3$N is $<$\,3.8. This means that even if HC$_3$HCN is not detected, it could still be more abundant than CH$_2$C$_3$N.
The nonobservation of the HC$_3$HCN radical can be explained by the large partition function for HC$_3$HCN compared to that for CH$_2$C$_3$N. For example, the partition function at 10\,K for HC$_3$HCN is 9462, while that for ortho-CH$_2$C$_3$N is 1718. Hence, the expected intensity of the lines of HC$_3$HCN, assuming the same column density as for CH$_2$C$_3$N, will be much lower, which makes its detection more difficult.
The chemistry of CH$_2$C$_3$N in TMC-1 has been discussed by \cite{Cabezas2021b}. The main formation reactions are C + CH$_2$CHCN, C$_2$ + CH$_3$CN, CN + CH$_2$CCH, and CH$_3$C$_3$NH$^+$ + e$^-$. The best-studied of these reactions is C + CH$_2$CHCN \citep{Su2005,Guo2006}. Theoretical calculations predict that in this reaction HC$_3$HCN is formed five times faster than CH$_2$C$_3$N \citep{Su2005}, while crossed-beam experiments also favor HC$_3$HCN as the main product, at least twice more than CH$_2$C$_3$N \citep{Guo2006}. It is still necessary to study the other potential routes to these radicals mentioned above. However, if the reaction C + CH$_2$CHCN is one of the dominant routes to CH$_2$C$_3$N in TMC-1, as suggested by the chemical model presented by \cite{Cabezas2021b}, and if this reaction produces HC$_3$HCN with an even higher branching ratio than CH$_2$C$_3$N, it is likely that a deeper integration will lead to the detection of the HC$_3$HCN radical in TMC-1. This will also allow us to better constrain the formation routes to these radicals and their role as intermediates in the buildup of large organic molecules in cold dark clouds.

\section{Conclusion}

We report the investigation of the rotational spectrum of 1-cyano propargyl radical, HC$_3$HCN. This transient species was generated in the gas phase using an electric discharges technique and was then characterized by Fourier transform microwave spectroscopy. A total of 193 hyperfine components that originated from 12 rotational transitions were observed and analyzed with a $^2A^{''}$ Hamiltonian. Accurate values for 22 molecular constants were derived from the analysis, indicating that the complicated hyperfine structures caused by the three nuclei with nonzero nuclear spin are well described by the employed Hamiltonian. The values of the hyperfine constants for the hydrogen and nitrogen nuclei were compared to those found for the CH$_2$C$_3$N isomer, and it seems that both radicals share some electronic similarities. Reliable frequency predictions were obtained from the molecular parameters derived from the spectroscopic analysis. These predictions were used to search for HC$_3$HCN in TMC-1 using the QUIJOTE line survey. We cannot confirm the presence of HC$_3$HCN in TMC-1, but we obtained a 3$\sigma$ upper limit to its column density of 6.0$\times$10$^{11}$ cm$^{-2}$.

\begin{acknowledgements}

The Spanish authors thank ERC for funding through grant ERC-2013-Syg-610256-NANOCOSMOS and Ministerio de Ciencia e Innovaci\'on for funding support through projects PID2019-106235GB-I00 and PID2019-107115GB-C21 / AEI / 10.13039/501100011033. MA thanks Ministerio de Ciencia e Innovaci\'on for grant RyC-2014-16277. Y. Endo thanks Ministry of Science and Technology of Taiwan through grant MOST108-2113-M-009-25.

\end{acknowledgements}

\begin{appendix}

\section{Laboratory observed transition frequencies for HC$_3$HCN}

\onecolumn
\begin{tiny}
\begin{longtable}{cccccccccccccccccc}
\caption[]{Observed transition frequencies for HC$_3$HCN.
\label{tab_lab_freq}}\\
\hline
\hline
 $N'$ & $K'_a$ & $K'_c$ & $J'$ & $F'_1$ & $F'_2$ & $F'$ & $N''$ & $K''_a$ & $K''_c$ & $J''$ & $F''_1$ & $F''_2$ & $F''$ & $\nu_{obs}$  &  $\nu_{calc}$  &   Obs-Calc & Int.$^a$ \\
      &        &        &      &        &        &      &      &        &        &      &        &        &      &   (MHz)      &   (MHz)        &   (MHz)    &      \\
\hline
\endfirsthead
\caption{continued.}\\
\hline
\hline
 $N'$ & $K'_a$ & $K'_c$ & $J'$ & $F'_1$ & $F'_2$ & $F'$ & $N''$ & $K''_a$ & $K''_c$ & $J''$ & $F''_1$ & $F''_2$ & $F''$ & $\nu_{obs}$  &  $\nu_{calc}$  &   Obs-Calc & Int.$^a$ \\
      &        &        &      &        &        &      &      &        &        &      &        &        &      &   (MHz)      &   (MHz)        &   (MHz)    &      \\
\hline
\endhead
\hline
\endfoot
\hline
\endlastfoot
\hline
 2 & 1 & 2 & 1.5 & 2.0 & 2.5 & 2.5 & 1 & 1 & 1 & 0.5 & 1.0 & 1.5 & 1.5 &  10470.944  & 10470.945  &  -0.001 &  1.368  \\
 2 & 1 & 2 & 1.5 & 2.0 & 2.5 & 3.5 & 1 & 1 & 1 & 0.5 & 1.0 & 1.5 & 2.5 &  10472.128  & 10472.128  &  -0.000 &  2.269  \\
 2 & 1 & 2 & 2.5 & 3.0 & 2.5 & 3.5 & 1 & 1 & 1 & 1.5 & 1.0 & 1.5 & 2.5 &  10479.894  & 10479.894  &  -0.000 &  1.420  \\
 2 & 1 & 2 & 2.5 & 3.0 & 3.5 & 3.5 & 1 & 1 & 1 & 1.5 & 2.0 & 2.5 & 2.5 &  10483.322  & 10483.322  &   0.000 &  2.123  \\
 2 & 1 & 2 & 2.5 & 3.0 & 3.5 & 4.5 & 1 & 1 & 1 & 1.5 & 2.0 & 2.5 & 3.5 &  10486.000  & 10486.001  &  -0.001 &  2.994  \\
 2 & 1 & 2 & 2.5 & 3.0 & 2.5 & 3.5 & 1 & 1 & 1 & 1.5 & 2.0 & 1.5 & 2.5 &  10491.230  & 10491.229  &   0.001 &  0.963  \\
 2 & 0 & 2 & 2.5 & 2.0 & 2.5 & 3.5 & 1 & 0 & 1 & 1.5 & 2.0 & 2.5 & 3.5 &  10721.774  & 10721.774  &  -0.000 &  1.069  \\
 2 & 0 & 2 & 1.5 & 1.0 & 1.5 & 1.5 & 1 & 0 & 1 & 0.5 & 0.0 & 0.5 & 0.5 &  10722.410  & 10722.410  &   0.000 &  0.623  \\
 2 & 0 & 2 & 1.5 & 1.0 & 1.5 & 2.5 & 1 & 0 & 1 & 0.5 & 1.0 & 1.5 & 1.5 &  10724.395  & 10724.393  &   0.002 &  1.509  \\
 2 & 0 & 2 & 1.5 & 2.0 & 1.5 & 2.5 & 1 & 0 & 1 & 0.5 & 0.0 & 0.5 & 1.5 &  10724.419  & 10724.421  &  -0.002 &  0.914  \\
 2 & 0 & 2 & 2.5 & 2.0 & 1.5 & 1.5 & 1 & 0 & 1 & 1.5 & 1.0 & 0.5 & 1.5 &  10725.576  & 10725.575  &   0.001 &  0.699  \\
 2 & 0 & 2 & 2.5 & 2.0 & 2.5 & 2.5 & 1 & 0 & 1 & 1.5 & 2.0 & 2.5 & 2.5 &  10726.048  & 10726.048  &   0.000 &  0.467  \\
 2 & 0 & 2 & 2.5 & 3.0 & 3.5 & 2.5 & 1 & 0 & 1 & 1.5 & 2.0 & 2.5 & 1.5 &  10726.750  & 10726.750  &   0.000 &  2.279  \\
 2 & 0 & 2 & 2.5 & 3.0 & 3.5 & 3.5 & 1 & 0 & 1 & 1.5 & 2.0 & 2.5 & 2.5 &  10727.055  & 10727.055  &   0.000 &  2.803  \\
 2 & 0 & 2 & 2.5 & 3.0 & 2.5 & 2.5 & 1 & 0 & 1 & 1.5 & 2.0 & 1.5 & 1.5 &  10727.110  & 10727.113  &  -0.003 &  2.013  \\
 2 & 0 & 2 & 2.5 & 3.0 & 2.5 & 3.5 & 1 & 0 & 1 & 1.5 & 2.0 & 1.5 & 2.5 &  10727.218  & 10727.218  &   0.000 &  3.175  \\
 2 & 0 & 2 & 2.5 & 3.0 & 3.5 & 4.5 & 1 & 0 & 1 & 1.5 & 2.0 & 2.5 & 3.5 &  10727.328  & 10727.326  &   0.002 &  4.000  \\
 2 & 0 & 2 & 2.5 & 2.0 & 1.5 & 2.5 & 1 & 0 & 1 & 1.5 & 1.0 & 0.5 & 1.5 &  10727.593  & 10727.593  &   0.000 &  1.657  \\
 2 & 0 & 2 & 1.5 & 2.0 & 2.5 & 2.5 & 1 & 0 & 1 & 1.5 & 1.0 & 1.5 & 1.5 &  10727.819  & 10727.818  &   0.001 &  2.331  \\
 2 & 0 & 2 & 2.5 & 2.0 & 2.5 & 2.5 & 1 & 0 & 1 & 0.5 & 1.0 & 0.5 & 1.5 &  10727.996  & 10727.995  &   0.001 &  1.611  \\
 2 & 0 & 2 & 1.5 & 2.0 & 2.5 & 3.5 & 1 & 0 & 1 & 1.5 & 1.0 & 1.5 & 2.5 &  10728.172  & 10728.173  &  -0.001 &  3.199  \\
 2 & 0 & 2 & 2.5 & 2.0 & 1.5 & 2.5 & 1 & 0 & 1 & 1.5 & 1.0 & 1.5 & 2.5 &  10728.324  & 10728.322  &   0.002 &  0.673  \\
 2 & 0 & 2 & 2.5 & 2.0 & 2.5 & 3.5 & 1 & 0 & 1 & 0.5 & 1.0 & 1.5 & 2.5 &  10728.517  & 10728.517  &   0.000 &  1.759  \\
 2 & 0 & 2 & 2.5 & 2.0 & 1.5 & 1.5 & 1 & 0 & 1 & 1.5 & 1.0 & 0.5 & 0.5 &  10728.804  & 10728.802  &   0.002 &  0.784  \\
 2 & 0 & 2 & 1.5 & 2.0 & 1.5 & 0.5 & 1 & 0 & 1 & 0.5 & 1.0 & 0.5 & 0.5 &  10729.240  & 10729.240  &   0.000 &  0.601  \\
 2 & 0 & 2 & 2.5 & 2.0 & 1.5 & 0.5 & 1 & 0 & 1 & 1.5 & 1.0 & 0.5 & 0.5 &  10729.491  & 10729.493  &  -0.002 &  0.524  \\
 2 & 0 & 2 & 1.5 & 2.0 & 1.5 & 2.5 & 1 & 0 & 1 & 0.5 & 1.0 & 1.5 & 2.5 &  10730.484  & 10730.484  &   0.000 &  1.045  \\
 2 & 0 & 2 & 1.5 & 2.0 & 1.5 & 1.5 & 1 & 0 & 1 & 0.5 & 1.0 & 0.5 & 1.5 &  10731.401  & 10731.399  &   0.002 &  0.485  \\
 2 & 0 & 2 & 1.5 & 1.0 & 0.5 & 1.5 & 1 & 0 & 1 & 0.5 & 0.0 & 0.5 & 1.5 &  10732.834  & 10732.837  &  -0.003 &  0.631  \\
 2 & 1 & 1 & 2.5 & 2.0 & 2.5 & 3.5 & 1 & 1 & 0 & 1.5 & 1.0 & 1.5 & 2.5 &  10970.762  & 10970.763  &  -0.001 &  1.362  \\
 2 & 1 & 1 & 1.5 & 2.0 & 2.5 & 3.5 & 1 & 1 & 0 & 0.5 & 1.0 & 1.5 & 2.5 &  10972.811  & 10972.812  &  -0.001 &  2.241  \\
 2 & 1 & 1 & 1.5 & 2.0 & 2.5 & 2.5 & 1 & 1 & 0 & 0.5 & 1.0 & 1.5 & 1.5 &  10973.887  & 10973.886  &   0.001 &  1.523  \\
 2 & 1 & 1 & 2.5 & 3.0 & 2.5 & 3.5 & 1 & 1 & 0 & 1.5 & 1.0 & 1.5 & 2.5 &  10979.876  & 10979.875  &   0.001 &  0.488  \\
 2 & 1 & 1 & 2.5 & 3.0 & 2.5 & 2.5 & 1 & 1 & 0 & 1.5 & 1.0 & 1.5 & 1.5 &  10981.327  & 10981.329  &  -0.002 &  0.905  \\
 2 & 1 & 1 & 2.5 & 3.0 & 3.5 & 2.5 & 1 & 1 & 0 & 1.5 & 2.0 & 2.5 & 1.5 &  10984.470  & 10984.462  &   0.008 &  1.706  \\
 2 & 1 & 1 & 2.5 & 3.0 & 3.5 & 3.5 & 1 & 1 & 0 & 1.5 & 2.0 & 2.5 & 2.5 &  10984.571  & 10984.572  &  -0.001 &  2.320  \\
 2 & 1 & 1 & 2.5 & 3.0 & 3.5 & 4.5 & 1 & 1 & 0 & 1.5 & 2.0 & 2.5 & 3.5 &  10985.861  & 10985.863  &  -0.002 &  2.997  \\
 2 & 1 & 1 & 2.5 & 3.0 & 2.5 & 3.5 & 1 & 1 & 0 & 1.5 & 2.0 & 1.5 & 2.5 &  10987.152  & 10987.152  &   0.000 &  1.892  \\
 2 & 1 & 1 & 2.5 & 3.0 & 2.5 & 2.5 & 1 & 1 & 0 & 1.5 & 2.0 & 1.5 & 1.5 &  10987.651  & 10987.652  &  -0.001 &  0.629  \\
 3 & 1 & 3 & 2.5 & 3.0 & 2.5 & 2.5 & 2 & 1 & 2 & 1.5 & 2.0 & 1.5 & 1.5 &  15712.568  & 15712.567  &   0.001 &  1.747  \\
 3 & 1 & 3 & 2.5 & 2.0 & 2.5 & 3.5 & 2 & 1 & 2 & 1.5 & 1.0 & 1.5 & 2.5 &  15712.920  & 15712.919  &   0.001 &  1.772  \\
 3 & 1 & 3 & 2.5 & 3.0 & 2.5 & 3.5 & 2 & 1 & 2 & 1.5 & 2.0 & 1.5 & 2.5 &  15713.285  & 15713.284  &   0.001 &  2.801  \\
 3 & 1 & 3 & 2.5 & 3.0 & 3.5 & 3.5 & 2 & 1 & 2 & 1.5 & 2.0 & 2.5 & 2.5 &  15715.069  & 15715.070  &  -0.001 &  2.741  \\
 3 & 1 & 3 & 3.5 & 3.0 & 3.5 & 4.5 & 2 & 1 & 2 & 2.5 & 2.0 & 2.5 & 3.5 &  15716.388  & 15716.388  &  -0.000 &  3.269  \\
 3 & 1 & 3 & 3.5 & 3.0 & 3.5 & 3.5 & 2 & 1 & 2 & 2.5 & 3.0 & 2.5 & 2.5 &  15716.600  & 15716.600  &  -0.000 &  2.422  \\
 3 & 1 & 3 & 3.5 & 3.0 & 2.5 & 3.5 & 2 & 1 & 2 & 2.5 & 2.0 & 1.5 & 2.5 &  15718.283  & 15718.281  &   0.002 &  2.859  \\
 3 & 1 & 3 & 3.5 & 4.0 & 4.5 & 3.5 & 2 & 1 & 2 & 2.5 & 3.0 & 3.5 & 2.5 &  15718.819  & 15718.819  &  -0.000 &  2.827  \\
 3 & 1 & 3 & 3.5 & 4.0 & 3.5 & 3.5 & 2 & 1 & 2 & 2.5 & 2.0 & 2.5 & 2.5 &  15719.133  & 15719.134  &  -0.001 &  2.259  \\
 3 & 1 & 3 & 3.5 & 4.0 & 4.5 & 4.5 & 2 & 1 & 2 & 2.5 & 3.0 & 3.5 & 3.5 &  15719.421  & 15719.421  &  -0.000 &  3.545  \\
 3 & 1 & 3 & 3.5 & 4.0 & 3.5 & 4.5 & 2 & 1 & 2 & 2.5 & 3.0 & 2.5 & 3.5 &  15719.921  & 15719.920  &   0.001 &  3.761  \\
 3 & 1 & 3 & 3.5 & 4.0 & 4.5 & 5.5 & 2 & 1 & 2 & 2.5 & 3.0 & 3.5 & 4.5 &  15720.414  & 15720.413  &   0.001 &  4.569  \\
 3 & 0 & 3 & 3.5 & 3.0 & 3.5 & 4.5 & 2 & 0 & 2 & 2.5 & 3.0 & 3.5 & 4.5 &  16081.880  & 16081.880  &   0.000 &  0.701  \\
 3 & 0 & 3 & 2.5 & 2.0 & 1.5 & 1.5 & 2 & 0 & 2 & 1.5 & 1.0 & 0.5 & 0.5 &  16084.905  & 16084.905  &  -0.000 &  0.580  \\
 3 & 0 & 3 & 3.5 & 3.0 & 2.5 & 2.5 & 2 & 0 & 2 & 2.5 & 2.0 & 1.5 & 2.5 &  16085.313  & 16085.314  &  -0.001 &  0.520  \\
 3 & 0 & 3 & 2.5 & 2.0 & 1.5 & 2.5 & 2 & 0 & 2 & 1.5 & 1.0 & 0.5 & 1.5 &  16085.596  & 16085.596  &   0.000 &  1.481  \\
 3 & 0 & 3 & 3.5 & 4.0 & 3.5 & 2.5 & 2 & 0 & 2 & 2.5 & 3.0 & 2.5 & 1.5 &  16086.044  & 16086.047  &  -0.003 &  2.182  \\
 3 & 0 & 3 & 3.5 & 4.0 & 4.5 & 3.5 & 2 & 0 & 2 & 2.5 & 3.0 & 3.5 & 2.5 &  16086.070  & 16086.069  &   0.001 &  3.315  \\
 3 & 0 & 3 & 3.5 & 4.0 & 4.5 & 4.5 & 2 & 0 & 2 & 2.5 & 3.0 & 3.5 & 3.5 &  16086.285  & 16086.282  &   0.003 &  4.191  \\
 3 & 0 & 3 & 3.5 & 4.0 & 3.5 & 4.5 & 2 & 0 & 2 & 2.5 & 3.0 & 2.5 & 3.5 &  16086.331  & 16086.331  &  -0.000 &  4.267  \\
 3 & 0 & 3 & 3.5 & 4.0 & 4.5 & 5.5 & 2 & 0 & 2 & 2.5 & 3.0 & 3.5 & 4.5 &  16086.393  & 16086.393  &   0.000 &  5.143  \\
 3 & 0 & 3 & 3.5 & 4.0 & 3.5 & 3.5 & 2 & 0 & 2 & 2.5 & 3.0 & 2.5 & 2.5 &  16086.507  & 16086.506  &   0.001 &  3.184  \\
 3 & 0 & 3 & 3.5 & 3.0 & 3.5 & 2.5 & 2 & 0 & 2 & 2.5 & 2.0 & 2.5 & 1.5 &  16086.787  & 16086.784  &   0.003 &  2.138  \\
 3 & 0 & 3 & 2.5 & 2.0 & 2.5 & 2.5 & 2 & 0 & 2 & 1.5 & 1.0 & 1.5 & 1.5 &  16086.820  & 16086.823  &  -0.003 &  1.912  \\
 3 & 0 & 3 & 2.5 & 3.0 & 2.5 & 3.5 & 2 & 0 & 2 & 1.5 & 2.0 & 1.5 & 2.5 &  16086.900  & 16086.898  &   0.002 &  2.326  \\
 3 & 0 & 3 & 3.5 & 3.0 & 3.5 & 3.5 & 2 & 0 & 2 & 2.5 & 2.0 & 2.5 & 2.5 &  16087.099  & 16087.098  &   0.001 &  2.828  \\
 3 & 0 & 3 & 3.5 & 3.0 & 2.5 & 3.5 & 2 & 0 & 2 & 2.5 & 2.0 & 1.5 & 2.5 &  16087.226  & 16087.227  &  -0.001 &  2.868  \\
 3 & 0 & 3 & 3.5 & 3.0 & 2.5 & 2.5 & 2 & 0 & 2 & 2.5 & 2.0 & 1.5 & 1.5 &  16087.331  & 16087.331  &  -0.000 &  2.019  \\
 3 & 0 & 3 & 3.5 & 3.0 & 3.5 & 4.5 & 2 & 0 & 2 & 2.5 & 2.0 & 2.5 & 3.5 &  16087.432  & 16087.432  &   0.000 &  3.523  \\
 3 & 0 & 3 & 2.5 & 3.0 & 2.5 & 2.5 & 2 & 0 & 2 & 1.5 & 2.0 & 1.5 & 1.5 &  16087.494  & 16087.497  &  -0.003 &  1.811  \\
 3 & 0 & 3 & 2.5 & 3.0 & 3.5 & 3.5 & 2 & 0 & 2 & 1.5 & 2.0 & 2.5 & 2.5 &  16087.505  & 16087.505  &   0.000 &  3.237  \\
 3 & 0 & 3 & 2.5 & 3.0 & 3.5 & 4.5 & 2 & 0 & 2 & 1.5 & 2.0 & 2.5 & 3.5 &  16087.651  & 16087.651  &  -0.000 &  4.284  \\
 3 & 0 & 3 & 2.5 & 2.0 & 2.5 & 3.5 & 2 & 0 & 2 & 1.5 & 1.0 & 1.5 & 2.5 &  16087.723  & 16087.723  &   0.000 &  2.988  \\
 3 & 0 & 3 & 2.5 & 3.0 & 3.5 & 2.5 & 2 & 0 & 2 & 1.5 & 2.0 & 2.5 & 1.5 &  16087.971  & 16087.970  &   0.001 &  2.279  \\
 3 & 0 & 3 & 3.5 & 3.0 & 2.5 & 1.5 & 2 & 0 & 2 & 2.5 & 2.0 & 1.5 & 1.5 &  16088.246  & 16088.249  &  -0.003 &  0.372  \\
 3 & 0 & 3 & 2.5 & 3.0 & 2.5 & 3.5 & 2 & 0 & 2 & 2.5 & 2.0 & 2.5 & 3.5 &  16088.864  & 16088.865  &  -0.001 &  0.869  \\
 3 & 0 & 3 & 2.5 & 2.0 & 1.5 & 0.5 & 2 & 0 & 2 & 1.5 & 1.0 & 0.5 & 0.5 &  16089.521  & 16089.518  &   0.003 &  0.618  \\
 3 & 0 & 3 & 2.5 & 3.0 & 2.5 & 2.5 & 2 & 0 & 2 & 2.5 & 2.0 & 2.5 & 2.5 &  16090.897  & 16090.901  &  -0.004 &  0.345  \\
 3 & 0 & 3 & 2.5 & 2.0 & 1.5 & 1.5 & 2 & 0 & 2 & 1.5 & 1.0 & 0.5 & 1.5 &  16091.982  & 16091.983  &  -0.001 &  0.725  \\
 3 & 1 & 2 & 2.5 & 2.0 & 2.5 & 3.5 & 2 & 1 & 1 & 1.5 & 1.0 & 1.5 & 2.5 &  16467.119  & 16467.119  &   0.000 &  2.692  \\
 3 & 1 & 2 & 2.5 & 2.0 & 1.5 & 2.5 & 2 & 1 & 1 & 1.5 & 1.0 & 0.5 & 1.5 &  16467.347  & 16467.346  &   0.001 &  1.426  \\
 3 & 1 & 2 & 3.5 & 3.0 & 3.5 & 4.5 & 2 & 1 & 1 & 2.5 & 2.0 & 2.5 & 3.5 &  16467.658  & 16467.659  &  -0.001 &  3.171  \\
 3 & 1 & 2 & 3.5 & 3.0 & 3.5 & 3.5 & 2 & 1 & 1 & 2.5 & 2.0 & 2.5 & 2.5 &  16468.581  & 16468.581  &  -0.000 &  2.440  \\
 3 & 1 & 2 & 3.5 & 4.0 & 3.5 & 2.5 & 2 & 1 & 1 & 2.5 & 3.0 & 2.5 & 1.5 &  16468.619  & 16468.614  &   0.005 &  1.600  \\
 3 & 1 & 2 & 2.5 & 2.0 & 2.5 & 2.5 & 2 & 1 & 1 & 1.5 & 1.0 & 1.5 & 1.5 &  16468.773  & 16468.773  &   0.000 &  1.709  \\
 3 & 1 & 2 & 2.5 & 3.0 & 3.5 & 4.5 & 2 & 1 & 1 & 1.5 & 2.0 & 2.5 & 3.5 &  16469.173  & 16469.173  &  -0.000 &  3.768  \\
 3 & 1 & 2 & 2.5 & 3.0 & 3.5 & 3.5 & 2 & 1 & 1 & 1.5 & 2.0 & 2.5 & 2.5 &  16469.273  & 16469.273  &  -0.000 &  2.856  \\
 3 & 1 & 2 & 2.5 & 3.0 & 2.5 & 3.5 & 2 & 1 & 1 & 2.5 & 3.0 & 2.5 & 2.5 &  16469.522  & 16469.520  &   0.002 &  2.323  \\
 3 & 1 & 2 & 2.5 & 3.0 & 3.5 & 2.5 & 2 & 1 & 1 & 1.5 & 2.0 & 2.5 & 1.5 &  16469.975  & 16469.975  &   0.000 &  2.066  \\
 3 & 1 & 2 & 3.5 & 3.0 & 2.5 & 2.5 & 2 & 1 & 1 & 2.5 & 2.0 & 1.5 & 1.5 &  16470.036  & 16470.035  &   0.001 &  1.727  \\
 3 & 1 & 2 & 3.5 & 3.0 & 2.5 & 3.5 & 2 & 1 & 1 & 2.5 & 2.0 & 1.5 & 2.5 &  16470.892  & 16470.892  &  -0.000 &  2.718  \\
 3 & 1 & 2 & 3.5 & 4.0 & 4.5 & 3.5 & 2 & 1 & 1 & 2.5 & 3.0 & 3.5 & 2.5 &  16471.097  & 16471.099  &  -0.002 &  2.976  \\
 3 & 1 & 2 & 3.5 & 4.0 & 4.5 & 4.5 & 2 & 1 & 1 & 2.5 & 3.0 & 3.5 & 3.5 &  16471.186  & 16471.185  &   0.001 &  3.746  \\
 3 & 1 & 2 & 3.5 & 4.0 & 3.5 & 4.5 & 2 & 1 & 1 & 2.5 & 3.0 & 2.5 & 3.5 &  16471.633  & 16471.636  &  -0.003 &  3.762  \\
 3 & 1 & 2 & 3.5 & 4.0 & 4.5 & 5.5 & 2 & 1 & 1 & 2.5 & 3.0 & 3.5 & 4.5 &  16471.658  & 16471.657  &   0.001 &  4.570  \\
 4 & 1 & 4 & 3.5 & 3.0 & 2.5 & 2.5 & 3 & 1 & 3 & 2.5 & 2.0 & 1.5 & 1.5 &  20951.308  & 20951.307  &   0.001 &  1.718  \\
 4 & 1 & 4 & 3.5 & 3.0 & 3.5 & 3.5 & 3 & 1 & 3 & 2.5 & 2.0 & 2.5 & 2.5 &  20951.588  & 20951.588  &   0.000 &  2.489  \\
 4 & 1 & 4 & 3.5 & 4.0 & 3.5 & 3.5 & 3 & 1 & 3 & 2.5 & 3.0 & 2.5 & 2.5 &  20952.247  & 20952.248  &  -0.001 &  2.925  \\
 4 & 1 & 4 & 3.5 & 4.0 & 3.5 & 4.5 & 3 & 1 & 3 & 2.5 & 3.0 & 2.5 & 3.5 &  20952.843  & 20952.841  &   0.002 &  3.992  \\
 4 & 1 & 4 & 3.5 & 3.0 & 3.5 & 4.5 & 3 & 1 & 3 & 2.5 & 2.0 & 2.5 & 3.5 &  20952.883  & 20952.883  &  -0.000 &  3.534  \\
 4 & 1 & 4 & 4.5 & 4.0 & 3.5 & 2.5 & 3 & 1 & 3 & 3.5 & 3.0 & 2.5 & 1.5 &  20953.203  & 20953.204  &  -0.001 &  2.033  \\
 4 & 1 & 4 & 3.5 & 3.0 & 2.5 & 3.5 & 3 & 1 & 3 & 2.5 & 2.0 & 1.5 & 2.5 &  20953.419  & 20953.418  &   0.001 &  2.549  \\
 4 & 1 & 4 & 3.5 & 4.0 & 4.5 & 4.5 & 3 & 1 & 3 & 2.5 & 3.0 & 3.5 & 3.5 &  20953.552  & 20953.552  &  -0.000 &  3.934  \\
 4 & 1 & 4 & 3.5 & 4.0 & 4.5 & 3.5 & 3 & 1 & 3 & 3.5 & 3.0 & 2.5 & 2.5 &  20953.712  & 20953.713  &  -0.001 &  3.080  \\
 4 & 1 & 4 & 4.5 & 4.0 & 4.5 & 3.5 & 3 & 1 & 3 & 3.5 & 3.0 & 3.5 & 2.5 &  20953.820  & 20953.820  &  -0.000 &  2.884  \\
 4 & 1 & 4 & 4.5 & 4.0 & 4.5 & 4.5 & 3 & 1 & 3 & 3.5 & 3.0 & 3.5 & 3.5 &  20953.946  & 20953.947  &  -0.001 &  3.683  \\
 4 & 1 & 4 & 3.5 & 4.0 & 4.5 & 5.5 & 3 & 1 & 3 & 2.5 & 3.0 & 3.5 & 4.5 &  20954.101  & 20954.100  &   0.001 &  4.996  \\
 4 & 1 & 4 & 4.5 & 4.0 & 4.5 & 5.5 & 3 & 1 & 3 & 3.5 & 3.0 & 3.5 & 4.5 &  20954.154  & 20954.154  &  -0.000 &  4.633  \\
 4 & 1 & 4 & 4.5 & 5.0 & 4.5 & 3.5 & 3 & 1 & 3 & 3.5 & 4.0 & 3.5 & 2.5 &  20954.830  & 20954.830  &  -0.000 &  2.631  \\
 4 & 1 & 4 & 4.5 & 4.0 & 3.5 & 4.5 & 3 & 1 & 3 & 3.5 & 3.0 & 2.5 & 3.5 &  20954.944  & 20954.943  &   0.001 &  4.028  \\
 4 & 1 & 4 & 4.5 & 5.0 & 5.5 & 4.5 & 3 & 1 & 3 & 3.5 & 4.0 & 4.5 & 3.5 &  20954.990  & 20954.991  &  -0.001 &  3.975  \\
 4 & 1 & 4 & 4.5 & 5.0 & 4.5 & 4.5 & 3 & 1 & 3 & 3.5 & 4.0 & 3.5 & 3.5 &  20955.306  & 20955.307  &  -0.001 &  3.952  \\
 4 & 1 & 4 & 4.5 & 5.0 & 5.5 & 5.5 & 3 & 1 & 3 & 3.5 & 4.0 & 4.5 & 4.5 &  20955.421  & 20955.421  &  -0.000 &  4.773  \\
 4 & 1 & 4 & 4.5 & 5.0 & 4.5 & 5.5 & 3 & 1 & 3 & 3.5 & 4.0 & 3.5 & 4.5 &  20955.736  & 20955.736  &   0.000 &  4.991  \\
 4 & 1 & 4 & 4.5 & 5.0 & 5.5 & 6.5 & 3 & 1 & 3 & 3.5 & 4.0 & 4.5 & 5.5 &  20955.881  & 20955.881  &  -0.000 &  5.832  \\
 4 & 0 & 4 & 4.5 & 4.0 & 3.5 & 3.5 & 3 & 0 & 3 & 3.5 & 3.0 & 2.5 & 3.5 &  21439.331  & 21439.332  &  -0.001 &  0.386  \\
 4 & 0 & 4 & 4.5 & 5.0 & 5.5 & 4.5 & 3 & 0 & 3 & 3.5 & 4.0 & 4.5 & 3.5 &  21439.986  & 21439.987  &  -0.001 &  4.337  \\
 4 & 0 & 4 & 4.5 & 5.0 & 4.5 & 3.5 & 3 & 0 & 3 & 3.5 & 4.0 & 3.5 & 2.5 &  21440.034  & 21440.034  &  -0.000 &  3.244  \\
 4 & 0 & 4 & 4.5 & 5.0 & 5.5 & 5.5 & 3 & 0 & 3 & 3.5 & 4.0 & 4.5 & 4.5 &  21440.168  & 21440.167  &   0.001 &  5.240  \\
 4 & 0 & 4 & 4.5 & 5.0 & 4.5 & 5.5 & 3 & 0 & 3 & 3.5 & 4.0 & 3.5 & 4.5 &  21440.194  & 21440.191  &   0.003 &  5.318  \\
 4 & 0 & 4 & 4.5 & 5.0 & 5.5 & 6.5 & 3 & 0 & 3 & 3.5 & 4.0 & 4.5 & 5.5 &  21440.231  & 21440.231  &   0.000 &  6.222  \\
 4 & 0 & 4 & 4.5 & 5.0 & 4.5 & 4.5 & 3 & 0 & 3 & 3.5 & 4.0 & 3.5 & 3.5 &  21440.317  & 21440.319  &  -0.002 &  4.239  \\
 4 & 0 & 4 & 4.5 & 4.0 & 4.5 & 3.5 & 3 & 0 & 3 & 3.5 & 3.0 & 3.5 & 2.5 &  21440.635  & 21440.638  &  -0.003 &  3.189  \\
 4 & 0 & 4 & 4.5 & 4.0 & 4.5 & 4.5 & 3 & 0 & 3 & 3.5 & 3.0 & 3.5 & 3.5 &  21440.908  & 21440.906  &   0.002 &  3.958  \\
 4 & 0 & 4 & 3.5 & 3.0 & 2.5 & 2.5 & 3 & 0 & 3 & 2.5 & 2.0 & 1.5 & 1.5 &  21441.044  & 21441.052  &  -0.008 &  1.795  \\
 4 & 0 & 4 & 4.5 & 4.0 & 3.5 & 4.5 & 3 & 0 & 3 & 3.5 & 3.0 & 2.5 & 3.5 &  21441.126  & 21441.124  &   0.002 &  4.070  \\
 4 & 0 & 4 & 4.5 & 4.0 & 4.5 & 5.5 & 3 & 0 & 3 & 3.5 & 3.0 & 3.5 & 4.5 &  21441.178  & 21441.176  &   0.002 &  4.786  \\
 4 & 0 & 4 & 4.5 & 4.0 & 3.5 & 3.5 & 3 & 0 & 3 & 3.5 & 3.0 & 2.5 & 2.5 &  21441.242  & 21441.246  &  -0.004 &  3.140  \\
 4 & 0 & 4 & 3.5 & 4.0 & 3.5 & 4.5 & 3 & 0 & 3 & 2.5 & 3.0 & 2.5 & 3.5 &  21441.323  & 21441.321  &   0.002 &  3.641  \\
 4 & 0 & 4 & 3.5 & 4.0 & 3.5 & 3.5 & 3 & 0 & 3 & 2.5 & 3.0 & 2.5 & 2.5 &  21441.455  & 21441.453  &   0.002 &  2.954  \\
 4 & 0 & 4 & 4.5 & 4.0 & 3.5 & 2.5 & 3 & 0 & 3 & 3.5 & 3.0 & 2.5 & 1.5 &  21441.484  & 21441.483  &   0.001 &  2.277  \\
 4 & 0 & 4 & 3.5 & 4.0 & 4.5 & 5.5 & 3 & 0 & 3 & 2.5 & 3.0 & 3.5 & 4.5 &  21441.670  & 21441.671  &  -0.001 &  5.331  \\
 4 & 0 & 4 & 3.5 & 4.0 & 4.5 & 4.5 & 3 & 0 & 3 & 2.5 & 3.0 & 3.5 & 3.5 &  21441.684  & 21441.683  &   0.001 &  4.254  \\
 4 & 0 & 4 & 3.5 & 3.0 & 3.5 & 3.5 & 3 & 0 & 3 & 2.5 & 2.0 & 2.5 & 2.5 &  21441.811  & 21441.811  &   0.000 &  3.075  \\
 4 & 0 & 4 & 3.5 & 4.0 & 4.5 & 3.5 & 3 & 0 & 3 & 2.5 & 3.0 & 3.5 & 2.5 &  21442.054  & 21442.058  &  -0.004 &  3.332  \\
 4 & 0 & 4 & 3.5 & 3.0 & 3.5 & 2.5 & 3 & 0 & 3 & 2.5 & 2.0 & 2.5 & 1.5 &  21442.089  & 21442.085  &   0.004 &  2.261  \\
 4 & 0 & 4 & 3.5 & 3.0 & 3.5 & 4.5 & 3 & 0 & 3 & 2.5 & 2.0 & 2.5 & 3.5 &  21442.130  & 21442.130  &  -0.000 &  4.147  \\
 4 & 0 & 4 & 3.5 & 4.0 & 3.5 & 4.5 & 3 & 0 & 3 & 3.5 & 3.0 & 3.5 & 4.5 &  21442.755  & 21442.755  &   0.000 &  0.649  \\
 4 & 1 & 3 & 4.5 & 4.0 & 4.5 & 5.5 & 3 & 1 & 2 & 3.5 & 3.0 & 3.5 & 4.5 &  21956.727  & 21956.726  &   0.001 &  4.510  \\
 4 & 1 & 3 & 4.5 & 4.0 & 3.5 & 3.5 & 3 & 1 & 2 & 3.5 & 4.0 & 3.5 & 2.5 &  21957.077  & 21957.076  &   0.001 &  2.859  \\
 4 & 1 & 3 & 4.5 & 4.0 & 4.5 & 4.5 & 3 & 1 & 2 & 3.5 & 3.0 & 3.5 & 3.5 &  21957.112  & 21957.110  &   0.002 &  3.673  \\
 4 & 1 & 3 & 4.5 & 4.0 & 4.5 & 3.5 & 3 & 1 & 2 & 3.5 & 3.0 & 3.5 & 2.5 &  21957.368  & 21957.369  &  -0.001 &  2.989  \\
 4 & 1 & 3 & 4.5 & 5.0 & 4.5 & 4.5 & 3 & 1 & 2 & 3.5 & 4.0 & 3.5 & 3.5 &  21957.455  & 21957.455  &   0.000 &  3.600  \\
 4 & 1 & 3 & 3.5 & 3.0 & 3.5 & 4.5 & 3 & 1 & 2 & 2.5 & 2.0 & 2.5 & 3.5 &  21957.508  & 21957.509  &  -0.001 &  3.909  \\
 4 & 1 & 3 & 4.5 & 5.0 & 5.5 & 4.5 & 3 & 1 & 2 & 3.5 & 4.0 & 4.5 & 3.5 &  21957.957  & 21957.959  &  -0.002 &  4.117  \\
 4 & 1 & 3 & 4.5 & 5.0 & 5.5 & 5.5 & 3 & 1 & 2 & 3.5 & 4.0 & 4.5 & 4.5 &  21957.999  & 21957.997  &   0.002 &  4.955  \\
 4 & 1 & 3 & 4.5 & 5.0 & 5.5 & 6.5 & 3 & 1 & 2 & 3.5 & 4.0 & 4.5 & 5.5 &  21958.183  & 21958.185  &  -0.002 &  5.833  \\
 4 & 1 & 3 & 4.5 & 5.0 & 4.5 & 5.5 & 3 & 1 & 2 & 3.5 & 4.0 & 3.5 & 4.5 &  21958.210  & 21958.204  &   0.006 &  4.969  \\
 4 & 1 & 3 & 3.5 & 4.0 & 4.5 & 5.5 & 3 & 1 & 2 & 2.5 & 3.0 & 3.5 & 4.5 &  21958.333  & 21958.334  &  -0.001 &  4.978  \\
 4 & 1 & 3 & 3.5 & 4.0 & 3.5 & 2.5 & 3 & 1 & 2 & 2.5 & 3.0 & 2.5 & 1.5 &  21958.482  & 21958.491  &  -0.009 &  2.135  \\
 4 & 1 & 3 & 3.5 & 4.0 & 4.5 & 3.5 & 3 & 1 & 2 & 2.5 & 3.0 & 3.5 & 2.5 &  21958.638  & 21958.637  &   0.001 &  3.153  \\
 4 & 1 & 3 & 3.5 & 4.0 & 3.5 & 4.5 & 3 & 1 & 2 & 3.5 & 3.0 & 2.5 & 3.5 &  21958.685  & 21958.683  &   0.002 &  3.926  \\
 4 & 1 & 3 & 3.5 & 3.0 & 3.5 & 2.5 & 3 & 1 & 2 & 2.5 & 2.0 & 2.5 & 1.5 &  21958.782  & 21958.782  &   0.000 &  2.118  \\
 1 & 1 & 0 & 1.5 & 2.0 & 1.5 & 2.5 & 1 & 0 & 1 & 1.5 & 2.0 & 1.5 & 2.5 &  26595.729  & 26595.728  &   0.001 &  1.559  \\
 1 & 1 & 0 & 1.5 & 2.0 & 2.5 & 3.5 & 1 & 0 & 1 & 1.5 & 2.0 & 2.5 & 3.5 &  26597.165  & 26597.163  &   0.002 &  2.577  \\
 1 & 1 & 0 & 1.5 & 1.0 & 1.5 & 2.5 & 1 & 0 & 1 & 0.5 & 1.0 & 1.5 & 1.5 &  26599.936  & 26599.939  &  -0.003 &  0.771  \\
 1 & 1 & 0 & 1.5 & 2.0 & 2.5 & 2.5 & 1 & 0 & 1 & 1.5 & 2.0 & 2.5 & 2.5 &  26601.798  & 26601.798  &   0.000 &  1.824  \\
 1 & 1 & 0 & 1.5 & 1.0 & 1.5 & 2.5 & 1 & 0 & 1 & 1.5 & 2.0 & 1.5 & 2.5 &  26603.007  & 26603.005  &   0.002 &  0.501  \\
 1 & 1 & 0 & 1.5 & 2.0 & 2.5 & 3.5 & 1 & 0 & 1 & 0.5 & 1.0 & 1.5 & 2.5 &  26603.908  & 26603.905  &   0.003 &  1.362  \\
 1 & 1 & 0 & 1.5 & 1.0 & 1.5 & 1.5 & 1 & 0 & 1 & 1.5 & 2.0 & 1.5 & 1.5 &  26604.708  & 26604.708  &  -0.000 &  0.751  \\
 1 & 1 & 0 & 1.5 & 1.0 & 1.5 & 2.5 & 1 & 0 & 1 & 1.5 & 2.0 & 2.5 & 3.5 &  26621.964  & 26621.962  &   0.002 &  0.753  \\
 1 & 1 & 0 & 0.5 & 1.0 & 1.5 & 1.5 & 1 & 0 & 1 & 1.5 & 1.0 & 1.5 & 1.5 &  26630.457  & 26630.458  &  -0.001 &  1.070  \\
 5 & 0 & 5 & 5.5 & 6.0 & 5.5 & 4.5 & 4 & 0 & 4 & 4.5 & 5.0 & 4.5 & 3.5 &  26786.229  & 26786.228  &   0.001 &  1.454  \\
 5 & 0 & 5 & 5.5 & 6.0 & 6.5 & 5.5 & 4 & 0 & 4 & 4.5 & 5.0 & 5.5 & 4.5 &  26786.687  & 26786.688  &  -0.001 &  5.354  \\
 5 & 0 & 5 & 5.5 & 6.0 & 6.5 & 6.5 & 4 & 0 & 4 & 4.5 & 5.0 & 5.5 & 5.5 &  26786.852  & 26786.851  &   0.001 &  6.272  \\
 5 & 0 & 5 & 5.5 & 6.0 & 5.5 & 6.5 & 4 & 0 & 4 & 4.5 & 5.0 & 4.5 & 5.5 &  26786.882  & 26786.882  &   0.000 &  6.352  \\
 5 & 0 & 5 & 5.5 & 6.0 & 6.5 & 7.5 & 4 & 0 & 4 & 4.5 & 5.0 & 5.5 & 6.5 &  26786.912  & 26786.912  &   0.000 &  7.272  \\
 5 & 0 & 5 & 5.5 & 6.0 & 5.5 & 5.5 & 4 & 0 & 4 & 4.5 & 5.0 & 4.5 & 4.5 &  26786.945  & 26786.944  &   0.001 &  5.278  \\
 5 & 0 & 5 & 5.5 & 5.0 & 5.5 & 4.5 & 4 & 0 & 4 & 4.5 & 4.0 & 4.5 & 3.5 &  26787.304  & 26787.305  &  -0.001 &  4.227  \\
 5 & 0 & 5 & 5.5 & 5.0 & 5.5 & 5.5 & 4 & 0 & 4 & 4.5 & 4.0 & 4.5 & 4.5 &  26787.545  & 26787.545  &  -0.000 &  5.049  \\
 5 & 0 & 5 & 5.5 & 5.0 & 5.5 & 6.5 & 4 & 0 & 4 & 4.5 & 4.0 & 4.5 & 5.5 &  26787.785  & 26787.784  &   0.001 &  5.947  \\
 5 & 0 & 5 & 5.5 & 5.0 & 4.5 & 5.5 & 4 & 0 & 4 & 4.5 & 4.0 & 3.5 & 4.5 &  26787.828  & 26787.828  &   0.000 &  5.212  \\
 5 & 0 & 5 & 5.5 & 5.0 & 4.5 & 4.5 & 4 & 0 & 4 & 4.5 & 4.0 & 3.5 & 3.5 &  26787.988  & 26787.989  &  -0.001 &  4.212  \\
 5 & 0 & 5 & 4.5 & 5.0 & 4.5 & 5.5 & 4 & 0 & 4 & 3.5 & 4.0 & 3.5 & 4.5 &  26788.171  & 26788.169  &   0.002 &  4.844  \\
 5 & 0 & 5 & 4.5 & 5.0 & 4.5 & 3.5 & 4 & 0 & 4 & 4.5 & 4.0 & 3.5 & 2.5 &  26788.237  & 26788.237  &  -0.000 &  3.322  \\
 5 & 0 & 5 & 4.5 & 5.0 & 5.5 & 6.5 & 4 & 0 & 4 & 3.5 & 4.0 & 4.5 & 5.5 &  26788.433  & 26788.434  &  -0.001 &  6.360  \\
 5 & 0 & 5 & 4.5 & 4.0 & 3.5 & 3.5 & 4 & 0 & 4 & 3.5 & 3.0 & 2.5 & 2.5 &  26788.458  & 26788.456  &   0.002 &  2.943  \\
 5 & 0 & 5 & 4.5 & 5.0 & 5.5 & 5.5 & 4 & 0 & 4 & 3.5 & 4.0 & 4.5 & 4.5 &  26788.532  & 26788.532  &  -0.000 &  5.295  \\
 5 & 0 & 5 & 4.5 & 5.0 & 5.5 & 4.5 & 4 & 0 & 4 & 3.5 & 4.0 & 4.5 & 3.5 &  26788.840  & 26788.844  &  -0.004 &  4.361  \\
 5 & 0 & 5 & 4.5 & 4.0 & 4.5 & 4.5 & 4 & 0 & 4 & 3.5 & 3.0 & 3.5 & 3.5 &  26788.865  & 26788.864  &   0.001 &  4.167  \\
 5 & 0 & 5 & 4.5 & 4.0 & 4.5 & 5.5 & 4 & 0 & 4 & 3.5 & 3.0 & 3.5 & 4.5 &  26788.986  & 26788.985  &   0.001 &  5.237  \\
 5 & 0 & 5 & 4.5 & 4.0 & 4.5 & 3.5 & 4 & 0 & 4 & 3.5 & 3.0 & 3.5 & 2.5 &  26789.084  & 26789.083  &   0.001 &  3.311  \\
 2 & 1 & 1 & 2.5 & 3.0 & 2.5 & 3.5 & 2 & 0 & 2 & 2.5 & 3.0 & 2.5 & 3.5 &  26855.660  & 26855.663  &  -0.003 &  3.416  \\
 2 & 1 & 1 & 2.5 & 3.0 & 3.5 & 4.5 & 2 & 0 & 2 & 2.5 & 3.0 & 3.5 & 4.5 &  26855.700  & 26855.700  &   0.000 &  4.105  \\
 2 & 1 & 1 & 2.5 & 3.0 & 2.5 & 2.5 & 2 & 0 & 2 & 2.5 & 3.0 & 2.5 & 2.5 &  26858.924  & 26858.924  &  -0.000 &  2.020  \\
 2 & 1 & 1 & 2.5 & 3.0 & 3.5 & 3.5 & 2 & 0 & 2 & 2.5 & 3.0 & 3.5 & 3.5 &  26859.314  & 26859.315  &  -0.001 &  3.308  \\
 2 & 1 & 1 & 2.5 & 3.0 & 3.5 & 4.5 & 2 & 0 & 2 & 2.5 & 2.0 & 2.5 & 3.5 &  26861.251  & 26861.252  &  -0.001 &  0.867  \\
 2 & 1 & 1 & 2.5 & 2.0 & 1.5 & 2.5 & 2 & 0 & 2 & 2.5 & 2.0 & 1.5 & 2.5 &  26867.205  & 26867.206  &  -0.001 &  1.473  \\
 2 & 1 & 1 & 2.5 & 2.0 & 2.5 & 2.5 & 2 & 0 & 2 & 2.5 & 2.0 & 2.5 & 2.5 &  26868.565  & 26868.566  &  -0.001 &  1.441  \\
 2 & 1 & 1 & 2.5 & 2.0 & 2.5 & 3.5 & 2 & 0 & 2 & 2.5 & 2.0 & 2.5 & 3.5 &  26870.950  & 26870.950  &  -0.000 &  1.781  \\
 2 & 1 & 1 & 1.5 & 2.0 & 2.5 & 1.5 & 2 & 0 & 2 & 1.5 & 2.0 & 2.5 & 1.5 &  26874.004  & 26874.003  &   0.001 &  1.553  \\
 2 & 1 & 1 & 1.5 & 2.0 & 2.5 & 2.5 & 2 & 0 & 2 & 1.5 & 2.0 & 2.5 & 2.5 &  26876.528  & 26876.526  &   0.002 &  2.116  \\
 2 & 1 & 1 & 1.5 & 2.0 & 2.5 & 3.5 & 2 & 0 & 2 & 1.5 & 2.0 & 2.5 & 3.5 &  26877.421  & 26877.421  &  -0.000 &  3.227  \\
 2 & 1 & 1 & 1.5 & 1.0 & 1.5 & 2.5 & 2 & 0 & 2 & 1.5 & 1.0 & 1.5 & 2.5 &  26891.143  & 26891.143  &   0.000 &  1.829  \\
\hline
\end{longtable}
\tablefoot{$^a$Calculated line strength factor.\\}
\end{tiny}
\twocolumn

\end{appendix}

\end{document}